\documentclass[twoside]{article}
\usepackage{fleqn,espcrc2} 

\usepackage{graphicx}

\newcommand{\AmS}{{\protect\the\textfont2
  A\kern-.1667em\lower.5ex\hbox{M}\kern-.125emS}}

\title{Effect of $\gamma$-irradiation on superconducting transition temperature and 
resistive transition in polycrystalline YBa$_{2}$Cu$_{3}$O$_{(7-\delta)}$}

	\author{B. I. Belevtsev\address{B. Verkin Institute for Low Temperature Physics
                        \& Engineering, Kharkov, 310164, Ukraine}, 
	I. V. Volchok\address{Kharkov State Technical University of Agriculture,
	              Kharkov, 310002, Ukraine},
	N. V. Dalakova$^a$, V. I. Dotsenko$^a$, L. G. Ivanchenko$^a$,\\    
	A. V. Kuznichenko\address{Kharkov State University, 
	Kharkov, 310077, Ukraine}, and I. I. Logvinov$^a$}
	
\begin{document}	

\begin{abstract}
A  bulk polycrystalline sample of YBa$_{2}$Cu$_{3}$O$_{(7-\delta)}$ 
($\delta \approx 0.1$) has been irradiated by $\gamma$-rays with $^{60}$Co 
source. Non-monotonic behavior of T$_{c}$ with increasing irradiation dose $\Phi$ (up to 
about 220 MR) is observed: $T_{c}$ decreases at low doses ($\Phi \leq 50$~MR) 
from initial value ($\approx 93$~K) by about 2 K and then rises, forming a
minimum. At higher doses ($\Phi \geq 120$~MR) T$_{c}$ goes down again. The 
temperature width of resistive transition increases rather sharply with dose 
below 75 MR and drops somewhat at higher dose. The results observed are 
discussed, taking into account the granular structure of sample studied and 
the influence of $\gamma$-rays on intergrain Josephson coupling.

\vspace{1pc}
\end{abstract}

\maketitle
The influence of crystal-lattice disorder on superconductivity is one of 
the key points in understanding fundamental properties of high-$T_{c}$ 
superconductors (HTSCs). To a good approximation, the two main types of 
disorder, which are essential for superconductivity, can be distinguished 
\cite{bel}. The first is microscopic disorder associated with perturbations of 
the crystal lattice on the atomic scale ({\it e.g.}, impurities, vacancies). It is 
responsible for electron localization and other phenomena which can affect the 
superconducting order parameter. The second type of disorder is associated with 
structural inhomogeneity of superconductors (granular structure, phase 
separation etc.). The disorder scale in this case is far larger than interatomic 
distances, and, hence, this disorder is called macroscopic. The macroscopic 
disorder affects mainly the superconducting phase coherence. In experimental 
studies it is desirable to separate the effects of these types of disorder. 
Ignoring this point could lead to serious errors in interpretation of results. 
\par
The disordering of HTSC can be produced with $\gamma$-rays. Attenuation 
distances of $\gamma$-ray with energies of a few MeV are of order of a few 
centimeters that enables one to investigate bulk samples. The known works in 
this field are mostly relevant to  polycrystalline
YBa$_{2}$Cu$_{3}$O$_{(7-\delta)}$ (YBCO) (see \cite{kato,kuts,vasek,oz} and
refs. therein). They are quite contradictory. In some of them \cite{kuts} 
no influence of $\gamma$-rays on $T_c$ and resistivity, $\rho$, was found up to 
dose $\Phi \simeq 1000$~MR; whereas in others \cite{kato,vasek,oz} a marked 
decrease in $T_c$ and increase in $\rho$ were observed at quite low doses. 
\par
In this report, we present a study of the  effect of $\gamma$-rays on $T_c$ and 
the superconducting transition in bulk polycrystalline sample (with grains about 
12 $\mu$m) of YBCO with $\delta \approx 0.1$. Irradiation was accomplished with 
a $^{60}$Co source at room temperature in air up to a dose 
$\Phi \approx 220$~MR. The temperature dependence $\rho (T)$ is found to be
linear above $T_c$ up to 300 K, which is the usual behavior for optimally doped 
YBCO. We have defined the experimental $T_c$ to be the temperature at which 
normal resistance is halved. Before irradiation $T_c$ was about 93~K. The 
temperature $T_{cz}$, at which the resistance goes to zero, was used as a second 
characteristic of the resistive transition. The difference in $T_{c}$ and $T_{cz}$, 
$\delta T_{c} = T_{c}-T_{cz}$, is a quite definite measure of the width of the 
resistive transition. Sometimes, the temperature $T_{cb}$ at the onset of the
superconducting transition is used for characterization of HTSCs. But this 
temperature can be evaluated with much less precision than $T_{c}$ or $T_{cz}$. 
\begin{figure}[h,scale=2]
\vskip -0.7 cm
\begin{center}\leavevmode
\includegraphics[width=0.950\linewidth]{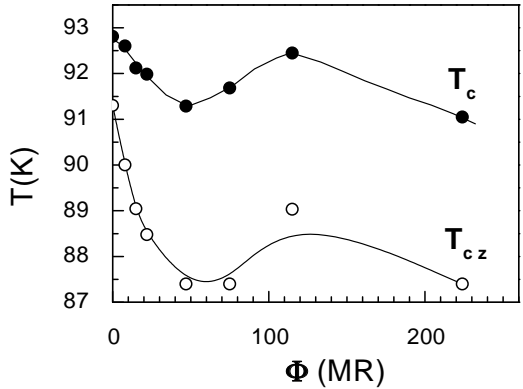}
\vskip -0.9 cm
\caption{The changes in $T_c$ and $T_{cz}$ with $\gamma$-ray dose. The solid
lines are guides to the eye.}
\label{fig2}
\end{center}
\end{figure}
\vskip -1.06cm 
\par
The changes in $T_c$ and $T_{cz}$ with $\gamma$-ray dose are shown in Fig.~1. It
can be seen that that $T_c$ decreases at low dose ($\Phi \leq 50$~MR) by 
$\approx 2$~K and then rises again, forming a minimum. At higher dose 
($\Phi \geq 120$~MR) $T_c$ clearly goes down again. The zero-resistance 
temperature $T_{cz}$ varies with $\gamma$-ray dose in a nearly same way as $T_c$, 
but with greater amplitude: the initial decrease is about 4 K. 
This means that $\delta T_c$ (which is a measure of sample inhomogeneity)
increases with dose up to $\Phi \approx 75$~MR. At higher doses $\delta T_c$ 
stops increasing and even drops somewhat. The effect of $\gamma$-rays on the 
resistivity was found only for temperatures close to $T_c$. No radiation effect in 
resistivity was detected above 200 K.
\par
Granularity of the sample should be taken into account in evaluating the 
results. The sample resistivity at room temperature ($\simeq 3.2$~m$\Omega$~cm)
is larger than that of the YBCO single-crystals by at least by a factor 10. 
The increased value comes from grain boundaries, which can be poorly conductive 
or even dielectric in HTSCs. At the same time the measured $T_c$   
($\approx 93$~K) corresponds to the highest $T_c$ in YBCO single crystals. This 
means the occurence of optimal current-carrying chains of grains with strong 
Josephson coupling.  
\par
Our calculations of the cross sections for displacement of lattice atoms in YBCO
by $\gamma$-rays due to the Compton process have shown that with commonly used 
$\gamma$-ray doses (up to 1000 MR) one should not expect any detectable 
variations in $\rho$ and $T_c$ in homogeneous crystals of YBCO. The effects 
observed in this work (as well as in the previous studies 
\cite{kato,vasek,oz}) are therefore undoubtedly connected with influence of 
$\gamma$-rays on the regions of grain boundaries. In HTSCs, the regions 
and their environtment are strongly depleted of charge carriers and 
thus can be very sensitive to $\gamma$-rays or particle irradiation.  
\par
The initial decrease in $T_c$ and $T_{cz}$ combined with the simultaneous 
increase in width of the resistive transition at low dose (Fig.~1) is quite 
expected for a percolating granular system. Optimal percolation current paths, 
which have ensured the high $T_c$ value before irradiation, surely have some 
``weak'' links. These are grain boundaries, which are strongly enough depleted with charge 
carriers and, therefore, are sensitive even to small radiation doses. 
Displacement of atoms from these areas can lead to carrier removal and, 
therefore, to deterioration of Josephson coupling in ``weak'' links. This can 
explain the observed decrease in $T_c$ and increase in 
$\delta T_{c}$ at low doses (Fig.~1). 
\par
Although the initial $T_c$ drop appears to be explicable, the general
non-monotonic picture in Fig.~1 is fairly surprising. To our knowledge, such
behavior has not been reported previously. It is likely that a second, 
independent mechanism of the $\gamma$-ray influence (maybe unrelated to 
radiation damage) operates concurrently with the above-mentioned one. This 
mechanism enhances the Josephson coupling and causes the increase in $T_c$ at 
higher doses. It can be connected with the ionizing influence of $\gamma$-rays. 
We will consider this hypothesis thoroughly in an extended paper.

\end{document}